\documentclass[reqno,11pt]{amsart}
\usepackage{hyperref}
\usepackage[mathscr]{eucal}
\usepackage{amsmath}
\usepackage{amssymb}
\usepackage{amsfonts}
\usepackage{amsthm}
\usepackage{graphicx}

\theoremstyle{definition}

\theoremstyle{definition}

\theoremstyle{definition}

\newcommand{\C}{\mathbb{C}}
\newcommand{\R}{\mathbb{R}}

\newcommand{\D}{\mathcal{D}}

\newcommand{\smallid}{\mbox{\rm{\scriptsize{1 \hspace{-1.05 em} 1}}}}

\def\Tr{{\rm Tr}}

\title[Quantum Oscillations in an Einstein-Dirac Cosmology]{Quantum Oscillations Can Prevent
the Big Bang Singularity in an Einstein-Dirac Cosmology}

\begin{document}

\author{Felix Finster}
\address{NWF I - Mathematik, Universit\"at Regensburg, D-93040 Regensburg, Germany}
\email{Felix.Finster@mathematik.uni-regensburg.de}
\thanks{F.F.\ is partially supported by the Deutsche Forschungsgemeinschaft. }

\author{Christian Hainzl}
\address{Departments of Mathematics and Physics, UAB, 1300 University
Blvd, Birmingham AL 35294, USA}
\email{hainzl@math.uab.edu}
\thanks{C.H.\ is partially supported by U.S. National Science
Foundation, grant DMS-0800906.}

\begin{abstract}
We consider a spatially homogeneous and isotropic system of Dirac particles coupled to
classical gravity. The dust and radiation dominated closed Friedmann-Robertson-Walker space-times
are recovered as limiting cases. We find a mechanism where quantum oscillations of the Dirac wave
functions can prevent the formation of the big bang or big crunch singularity. Thus before the big crunch,
the collapse of the universe is stopped by quantum effects and reversed to an expansion,
so that the universe opens up entering a new era of classical behavior.

Numerical examples of such space-times are given, and the dependence on various parameters is
discussed. Generically, one has a collapse after a finite number of cycles.
By fine-tuning the parameters we construct an example of a space-time which
is time-periodic, thus running through an infinite
number of contraction and expansion cycles.
\end{abstract}

\maketitle

\section{Introduction, the Einstein-Dirac System}
Near the big bang or big crunch singularity, the matter density and
the space-time curvature become arbitrarily large. It is generally
believed that in this regime, quantum effects should come into play,
which could even prevent the formation of the singularity. This
effect has first been analzyed in~\cite{parker+fulling} for
classical gravity coupled to a second quantized matter field; see
also~\cite{toporensky}. For a quantized gravitational field, this
effect has been studied in~\cite{padmanabhan}, and it was worked out
in more detail in the framework of string cosmology~\cite{TPS} and
in loop quantum gravity~\cite{bojowald}. In contrast to the above
approaches, we are here more modest and work with classical gravity
coupled to Dirac wave functions, without using second-quantized
fields. We find a new mechanism, based on the oscillations of the
spin of the matter field, which tends to prevent the formation of a
space-time singularity. The mechanism can be understood as
a {\em spin condensation} effect. An
advantage of our model is that it is very simple and can be analyzed
without any approximations. The Dirac field could be interpreted
physically as a new fermionic particle which becomes significant on
the cosmological scale. Alternatively, the Dirac field can be
understood similar to a ``quintessence''~\cite{wetterich} as
effectively describing unknown fields acting in our universe on the
large.
Describing such a cosmological field by the Dirac equation seems physically
natural and changes the behavior of the cosmological model drastically near the big
bang or big crunch, where quantum oscillations can prevent the formation of space-
time singularities depending on the initial conditions. Such a bounce back may happen
several times, before, generically, the solution collapses.
For brevity we here restrict attention to the case of
a closed universe. However, the flat and hyperbolic cases are less
interesting, since they expand infinitely. We remark
that the hyperbolic case allows for solutions with
negative energy density where the matter turns to be repulsive
(see~\cite{FH}).

For the derivation of the equations we follow the standard approach
in~\cite{penrose+rindler, U22} where we work with a torsion free
connection (for the effect of torsion see the cosmological model in
Einstein-Cartan theory~\cite{demianski}). The Einstein-Dirac (ED)
equations read
\begin{equation}\label{Einst-Dir}
R^i_{j} -\frac 12 R\:\delta^i_j = 8\pi \kappa\,T^i_{j}\:, \qquad (\D
- m) \Psi = 0\:,
\end{equation}
where $T^i_j$ is the energy-momentum tensor of the Dirac particles, $\kappa$ is the
gravitational constant, $\D$ denotes the Dirac operator,
and $\Psi$ is the Dirac wave function (we always work in natural units $\hbar=c=1$).
For the metric we take the ansatz of the closed Friedmann-Robertson-Walker (FRW) geometry
\[ ds^2 = dt^2 - R^2(t) \,d\sigma^2 , \]
where~$t$ is the time for an observer at rest, $R$ is the scale function, and~$d\sigma$ is the
line element on the unit $3$-sphere,
\[ d\sigma^2 = \frac {dr^2}{1 - r^2} + r^2 d\vartheta^2 + r^2 \sin^2 \vartheta\, d\varphi^2 \:, \]
where~$r \in (-1,1)$ is a radial variable, and~$(\vartheta, \varphi)
\in (0, \frac{\pi}{2}) \times [0, 2 \pi)$ are the angular variables.
The Dirac operator in this metric can be written as (see~\cite{U22})
\begin{equation}\label{Dequ} \D = i \gamma^0\left(\partial_t + \frac {3 \dot
R(t)}{2R(t)}\right) + \frac 1{R(t)} \left(\begin{matrix}
0 & \D_{S^3} \\ -\D_{S^3} & 0 \\
\end{matrix}\right) , \end{equation}
where $\gamma^0 = \left( \begin{smallmatrix} \smallid &0 \\ 0 & -\smallid \\ \end{smallmatrix} \right)$
is the usual $4\times 4$ Dirac matrix, and $\D_{S^3}$ is the Dirac operator on the
unit $3$-sphere. The operator~$\D_{S^3}$ has discrete eigenvalues
$\lambda = \pm \frac{3}{2}, \pm \frac{5}{2},\ldots$, corresponding to a quantization
of the possible momenta of the Dirac particles. For a given eigenvector~$\psi_\lambda$,
we can separate the Dirac equation with the ansatz
\begin{equation} \label{ansatz}
\Psi_\lambda = R(t)^{-\frac{3}{2}} \left[ \frac{8 \pi \kappa}{3}
\left(\lambda^2- \frac{1}{4} \right) \right]^{-\frac{1}{2}}
\left( \begin{matrix} \alpha(t) \:\psi_{\lambda}(r,\vartheta,\varphi) \\
\beta(t) \:\psi_\lambda(r,\vartheta,\varphi) \end{matrix} \right) ,
\end{equation}
thus describing the time dependence of the wave function by two
complex functions~$\alpha$ and~$\beta$. In order to be consistent
with the homogeneous and isotropic ansatz of the metric, the Dirac
spinors must also be in a homogeneous and isotropic configuration.
In analogy to the method employed in~\cite{ED} for spherical
symmetry, this can be achieved by taking an anti-symmetrized product
of the wave functions~(\ref{ansatz}), where~$\psi_{\lambda}$ runs
over an orthonormal basis of the eigenspace corresponding to a
fixed~$\lambda$, where the symmetry is expressed by the fact that
$\sum |\psi_\lambda(r,\vartheta,\phi)|^2 = \lambda^2 -\frac{1}{4}$,
with the sum running over all corresponding angular quantum numbers.
We thus obtain a fermionic Hartree-Fock state composed
of~$\lambda^2-\frac{1}{4}$ particles. The energy-momentum tensor
$T^i_j$ can be derived in a similar way as in \cite{ED}, see Equ.
(4.4). By spherical symmetry the components $T^t_r, T^t_\vartheta,
T^t_\phi $ have to vanish as well as the components $T^r_\phi,
T^r_\vartheta, T^\vartheta_\phi $, which is due to the homogeneity
in space. Therefore $T^i_j$ is a diagonal matrix where the space
components are equal, again due to the fact that our system is
homogeneous in space. Analogously to \cite{ED} we derive the time
component of the energy-momentum tensor, which reads
\[ T^t_t = \sum\Tr_{\C^4} \bigg\{  i \gamma^0\left(\partial_t + \frac {3 \dot R(t)}{2R(t)}\right)
|\Psi_\lambda\rangle \langle \Psi_\lambda | \bigg\} ,
\] where the sum runs over the angular momentum quantum numbers,
whose indices are suppressed for notational reason. Using
\eqref{ansatz} and the fact that the sum of
$|\psi_\lambda(r,\vartheta,\phi)|^2$ is constant in space, since the
sum runs over the angular quantum numbers corresponding to
$\lambda$, we obtain that
\begin{equation}\label{Ttt} 8 \pi \kappa T^t_t = \left[ m \Big(|\alpha|^2-|\beta|^2 \Big)
- \frac{2 \lambda}{R}\: \text{Re}(\alpha \overline{\beta}) \right].
\end{equation}
Substituting the ansatz~\eqref{ansatz} and \eqref{Ttt} into the ED
equations~\eqref{Einst-Dir}, we obtain the following system of ODEs
in the spinors~$(\alpha, \beta)$ and the scale function~$R$ (for a
detailed derivation see~\cite{FH}),
\begin{eqnarray}
i\frac{d}{dt} \left( \begin{matrix} \alpha  \\  \beta \end{matrix} \right)
&=&  \left( \begin{matrix} m & -\lambda/R \\ -\lambda/R & - m \end{matrix} \right) \label{dirac}
\left(\begin{matrix} \alpha \\  \beta \end{matrix}\right) \\
\dot{R}^2 + 1 &=& \frac{m}{R} \left( |\alpha|^2 - |\beta|^2 \right) -
\frac{\lambda}{R^2} \left(\overline{\beta} \alpha  +
\overline{\alpha} \beta \right) \label{einstein} .
\label{einstein2}
\end{eqnarray}
Let us mention that the equations obtained from the spacial
components of the energy-momentum tensor are automatically satisfied
due to the continuity equation, for which reason we omitted its
exact form. Let us emphasize again that  all particles in our model
have the same momentum $\lambda$ and therefore give rise to the same
spinor equation \eqref{dirac}, meaning they are represented by the
same spinor $(\alpha,\beta)$. This is what we mean by the
above-mentioned notion of {\em spin condensation}.

Note that~(\ref{einstein}) determines~$\dot{R}$ only up to a sign,
and at first sight this seems to lead to an ambiguity
whenever~$\dot{R}$ becomes zero. However, $\dot{R}$ becomes uniquely
determined by demanding that~$R$ be twice differentiable, implying
that~$\dot{R}$ must change sign at every zero of~$\dot{R}$.
Normalizing the probability integral to one, we obtain the condition
\begin{equation} \label{norm1}
|\alpha|^2 + |\beta|^2 = \frac{8 \pi \kappa}{3} \left(\lambda^2- \frac{1}{4} \right) .
\end{equation}
Differentiating and substituting~(\ref{dirac}), one sees that this normalization condition is indeed
time independent.

A short calculation shows that the equations~(\ref{dirac}, \ref{einstein}) are for any~$\rho>0$ invariant under the scalings
\[ R \rightarrow \rho R\:,\quad t \rightarrow \rho t \:,\quad
m \rightarrow \frac{m}{\rho}\:,\quad \lambda \rightarrow \lambda\:,\quad
(\alpha, \beta) \rightarrow \rho\, (\alpha, \beta)\:. \]
Since this scaling changes the norm of the spinors, it is no loss of generality to replace~(\ref{norm1})
by the simpler condition
\begin{equation} \label{normalization}
|\alpha|^2 + |\beta|^2 = 1\:.
\end{equation}
An other way of understanding the above rescaling is that we choose the units for the
gravitational constant such that the right side of~(\ref{norm1}) equals one.

The resulting ED system involves the physical parameters~$m$ (the rest mass of the Dirac particles)
and~$\lambda$ (the Dirac momentum, also related to the number of particles), as well as two free parameters to set the initial conditions of the spinors~$\alpha$ and~$\beta$. These four parameters
describe all the physical configurations of our system.

\section{The Bloch Representation}
For the analysis of the system of ODEs~(\ref{dirac}, \ref{einstein}) it is convenient to
regard the spinor $(\alpha, \beta)$ as a two-level quantum state, and to represent it by a Bloch
vector~$\vec{v}$. More precisely, introducing the $3$-vectors
\[ \vec{v} = \left\langle \!\left( \!\!\begin{array}{c} \alpha \\ \beta \end{array} \!\!\right)\!,
\vec{\sigma} \left( \!\!\begin{array}{c} \alpha \\ \beta \end{array} \!\!\right) \!\right\rangle_{\C^2}
\quad {\mbox{and}} \quad
\vec{b} = \frac{2 \lambda}{R}\, e_1 - 2 m e_3 \]
(where~$\vec{\sigma}$ are the Pauli matrices, and~$e_1, e_2, e_3$ are the standard basis
vectors in~$\R^3$), the ED equations become
\[ \dot{\vec{v}} = \vec{b} \wedge \vec{v} \:,\qquad
\dot{R}^2+1 = -\frac{1}{2R} \: \vec{b} \cdot \vec{v} \]
(where~`$\wedge$' and~`$\cdot$' denote the cross product and the
scalar product in Euclidean~$\R^3$, respectively). To further
simplify the equations, we introduce a rotation~$U$ around the
$e_2$-axis, such that~$\vec{b}$ becomes parallel to~$e_1$,
\[ U \vec{b} = \frac{2}{R}\: \sqrt{\lambda^2 + m^2 R^2}\, e_1\:. \]
Then the vector~$\vec{w} := U v$ satisfies the equations
\begin{equation} \label{bloch}
\boxed{ \quad \dot{\vec{w}} = \vec{d} \wedge \vec{w} \:,\qquad
\dot{R}^2+1 = -\frac{1}{R^2} \: \sqrt{\lambda^2 + m^2 R^2}\: w_1\:, \quad }
\end{equation}
where
\begin{equation} \label{ddef}
\vec{d} := \frac{2}{R}\: \sqrt{\lambda^2 + m^2 R^2}\: e_1 \:-\:
 \frac{\lambda m R}{\lambda^2 + m^2 R^2}\: \frac{\dot{R}}{R}\: e_2 \:.
\end{equation}
We refer to the two equations in~(\ref{bloch}) as the Dirac and Einstein equations in the Bloch representation, respectively. Notice that equation \eqref{ddef} can only have solutions if $w_1$ is negative.

In the Bloch representation, one can most easily recover the dust and radiation dominated
FRW-geometries. In the two limiting cases~$m \rightarrow 0$ and~$\lambda \rightarrow 0$, the
second term in~(\ref{ddef}) drops out, and thus the ED equations simplify respectively to
\begin{align}
 \dot{\vec{w}} &= \frac{2 |\lambda|}{R}\: e_1 \wedge \vec{w} \:,\qquad &
\dot{R}^2+1 &= -\frac{|\lambda|}{R^2} \: w_1 && \text{(radiation dominated universe)} \\
 \dot{\vec{w}} &= 2m\: e_1 \wedge \vec{w} \:,\qquad &
\dot{R}^2+1 &= -\frac{m}{R} \: w_1 && \text{(dust universe)}\:.
\end{align}
In both cases, the Dirac equation describes a rotation around the fixed axis~$e_1$. This implies
that~$w_1$ is constant in time, and thus the Einstein equations reduce to the well-known
Friedmann equations for a radiation dominated universe and a dust universe, respectively.
Note that the components~$w_2$ and~$w_3$ do oscillate due to quantum effects.
These oscillations can be understood as the familiar ``Zitterbewegung'' of Dirac particles.
But these quantum oscillations do not enter the Einstein equations.

Away from the above limiting cases, the function~$w_1$ is in general not
constant, but it takes part in the oscillations as described by the first equation in~\eqref{bloch}.
As a consequence, the dynamics can no longer be described by a single Friedmann equation,
but only by the coupled system of Einstein-Dirac equations~\eqref{bloch}.
The relevant length scale is characterized by the radius
\begin{equation} \label{Rqudef}
R_\text{qu} = \frac{\lambda}{m}\:.
\end{equation}
On this length scale, the quantum oscillations do enter the Einstein equations, leading
to effects which go beyond the scope of classical cosmology.
We point out that the radius~$R_\text{qu}$ can be much larger than the Planck length.
Thus, in contrast to~\cite{bojowald, TPS}, our mechanism is not related to quantum gravity;
instead we are working consistently within the framework of Einstein's general theory of relativity.

\section{Quantum Oscillations Preventing the Big Crunch}
\begin{figure}
\begin{center}
\includegraphics[width=6cm]{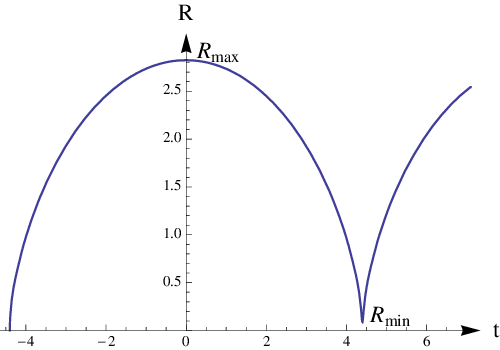} $\quad$
\includegraphics[width=6cm]{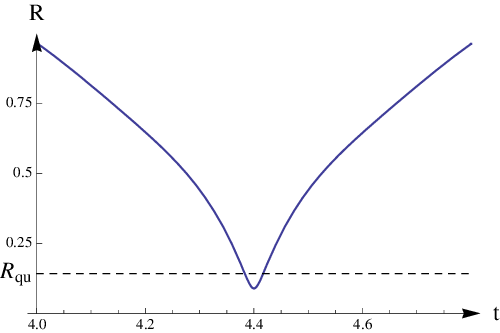} \\[1em]
\includegraphics[width=6cm]{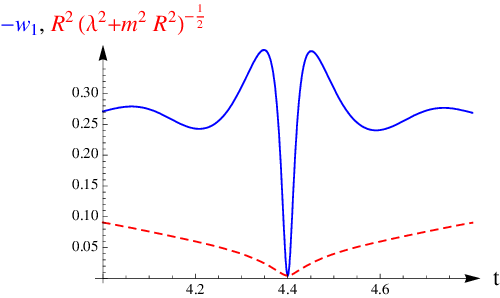} $\quad$
\includegraphics[width=6cm]{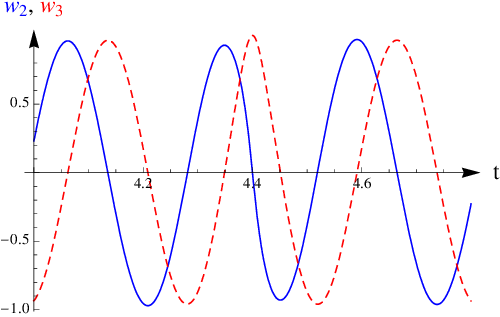}
\caption{Quantum oscillations preventing the big crunch for
$\lambda=\frac{3}{2}$, $m=10.5448$ and~$w_1(R_{\text{max}})=-0.2675$.}
\label{figcrunch}
\end{center}
\end{figure}

Using a standard ODE solver, we shoot for numerical solutions
starting at the point where the scale function~$R$ reaches its maximum
$R_\text{max}$, solving forward and backwards in time, until we reach
a big crunch or big bang singularity, respectively.
In~Figure~\ref{figcrunch} a typical solution is shown. The function~$R$
increases after the big bang similar as in the classical
FRW solution up to its maximal value~$R_\text{max}$, where it starts
decreasing.
The classical contraction stops on the scale~$R \sim R_\text{qu}$, where the quantum oscillations
become important. This leads to an oscillatory behavior of~$R$. As a consequence, the sign
of~$\dot{R}$ may flip at some minimal radius~$R_\text{min}$, where the universe
again begins to expand, going over to a classical FRW-like space-time.
The functions~$w_2$ and~$w_3$ oscillate, with slightly varying amplitude.
The function~$w_1$ is nearly constant in the classical region, but becomes oscillatory
near the quantum regime~$R \sim R_\text{qu}$ (see Figure~\ref{figcrunch}); its nonlinear
interplay with the scale function~$R$ leads to the surprising effect that the big crunch is
avoided.

\begin{figure}
\begin{center}
\includegraphics[width=8cm]{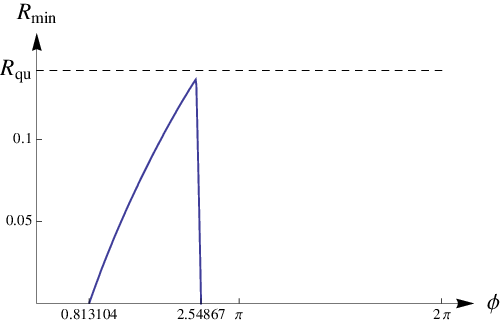}
\caption{Dependence of the minimal radius on the initial angle~$\phi$
for~$\lambda=\frac{3}{2}$, $m=10.5448$ and~$w_1(R_{\text{max}})=-0.2675$.}
\label{figprob}
\end{center}
\end{figure}

The behavior in the quantum regime depends crucially on the phase~$\phi$
of the Dirac oscillations, defined by~$\phi = \arctan(w_2/w_3)$.
In Figure~\ref{figprob} the minimal radius~$R_\text{min}$ is
plotted as a function of the initial value of~$\phi$.
If~$R_\text{min}>0$, the big crunch is avoided and~$R$ ``bounces back,'' whereas in the
case~$R_\text{min}=0$ the big crunch appears despite the quantum
effects. Considering the phase~$\phi$ as unknown, one can give the
result of Figure~\ref{figprob} a statistical interpretation: For a
random initial phase, this bounce appears with a finite
probability (in the example of Figure~\ref{figprob}, this
probability has the value $p=0.276224$).

\begin{figure}
\begin{center}
\includegraphics[width=6cm]{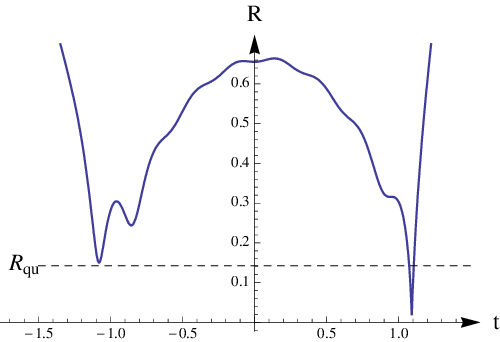} $\quad$
\includegraphics[width=6cm]{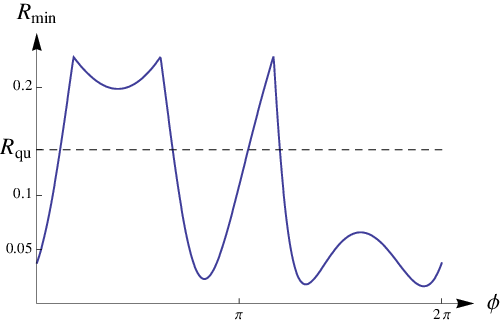}
\caption{Quantum oscillations preventing the big crunch for
$\lambda=\frac{3}{2}$, $m=10.5448$ and~$w_1(R_{\text{max}})=-0.0608$.}
\label{fignocrunch}
\end{center}
\end{figure}

Let us briefly discuss the qualitative dependence of the bounce on the free parameters~$R_\text{max}$,
$R_\text{qu}$, $\lambda$, and~$\phi$ of our model (for a detailed analysis we refer to~\cite{FH}).
First of all, the probability of preventing the crunch can be increased by choosing~$|w_1(R_\text{max})|$
smaller. This is shown in the example Figure~\ref{fignocrunch}, where the bounce occurs even with
probability one.
We point out that small~$|w_1(R_\text{max})|$ does not imply that~$R_\text{max}$
is also small. Namely, as one sees from the Einstein equation in~(\ref{bloch}) by setting~$\dot{R}$
equal to zero, a small value of~$|w_1(R_\text{max})|$ can be compensated by
choosing~$m$ or~$\lambda$ large. This leaves us the freedom to choose~$R_\text{qu}$, (\ref{Rqudef}),
at will. In this way, one can construct space-times with a high probability of bouncing
in the physically relevant case where~$R_\text{max}$ is the size of our universe
and~$R_\text{qu}$ is a microscopic length scale.

An interesting limiting case of our ED equations is obtained by dropping the first
summand in~(\ref{ddef}). This is a good approximation in the so-called {\em{scale dominated region}}
where~$R$ decreases so fast that the oscillations of~$w_2$ and~$w_3$ can be neglected.
In particular, the scale dominated region describes the
bounce in the physically interesting limit~$R_\text{max} \rightarrow \infty$ (for fixed~$\lambda$).
In this limiting case, we can integrate the Dirac equation in~(\ref{bloch}) explicitly
and express~$\vec w$  as a function of $R$,
\[ \vec{w}(t) = \begin{pmatrix} \cos \nu &0 & -\sin \nu \\ 0  &0&0 \\ \sin \nu &0& \cos \nu \end{pmatrix} \vec{w}(t_0) \:, \]
where $\nu(t) = -\arctan[R(t)/R_{\rm qu}] + \arctan[R(t_0)/R_{\rm qu}]$, and~$t_0$ is the
time from which on the scale dominated approximation applies.
For the probability~$p$ that the crunch is avoided one obtains
\[ p = \frac{1}{\pi}\: \arccos \left( \frac{|w_1(R_\text{max})|}{\sqrt{1-w_1(R_\text{max})^2}}\:
\frac{R_\text{qu}}{R_\text{max}} \right) . \]
This probability tends to~$1/2$ as~$R_{\rm max}$ tends to infinity, showing that
the bounce is of significance even for fixed~$|w_1(R_\text{max})|$ in an arbitrarily large universe.

\section{Time-Periodic Solutions}
\begin{figure}
\begin{center}
\includegraphics[width=6.5cm]{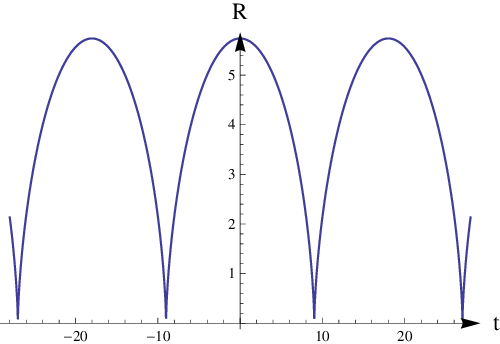}
\includegraphics[width=7.5cm]{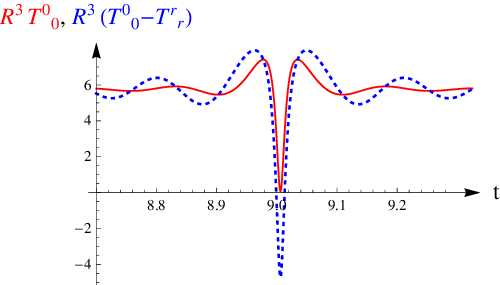}
\caption{A time-periodic ED cosmology for~$\lambda=\frac{3}{2}$,
$m=21.4286$, $\vec{w}_{|t=0} = (-0.267774, 0, -0.963482)$.}
\label{figperiod}
\end{center}
\end{figure}

By iteratively adjusting the starting values at~$t=0$, we
constructed time-periodic solutions. In Figure~\ref{figperiod} an
example of a time-periodic solution is shown. Except near the
``quantum turning points'' between collapse and expansion, the
space-time is well-approximated by the classical dust-dominated FRW
universe. The energy conditions are satisfied except near the
quantum turning points, as is shown on the
right of Figure~\ref{figperiod}. We point out that the periodic
solutions were constructed by fine-tuning the initial condition
and are thus not generic. A generic solution collapses after
a finite number of cycles.

\section{Conclusion}
A new class of solutions of the Einstein equations coupled to Dirac
spinors was constructed. These solutions satisfy the energy
conditions except at the quantum turning points and are thus physically relevant.
Our model reveals in a
simple and explicit setting a general mechanism which tends to avoid
space-time singularities, such as the big bang or the big crunch, if
the quantum mechanical nature of matter is taken into account. By
fine-tuning the initial conditions our model even allows for
periodic solutions, an eternal universe with an infinite number of
cycles.

\bigskip \noindent {\it Acknowledgements.}
We thank the referees for helpful comments.
We are grateful to the Vielberth Foundation, Regensburg, for generous support.
C.H.\ would like to thank the Erwin Schr\"odinger Institute, Vienna.


\providecommand{\bysame}{\leavevmode\hbox to3em{\hrulefill}\thinspace}
\providecommand{\MR}{\relax\ifhmode\unskip\space\fi MR }
\providecommand{\MRhref}[2]{%
  \href{http://www.ams.org/mathscinet-getitem?mr=#1}{#2}
}
\providecommand{\href}[2]{#2}

\end{document}